\documentclass{aastex}
\usepackage{spr-astr-addons}
\usepackage{url}

\RequirePackage{color}

\begin{document}

\title{Imbalanced magnetohydrodynamic turbulence modified by velocity shear
in the solar wind}
\slugcomment{Not to appear in Nonlearned J., 45.}
\shorttitle{Short article title}
\shortauthors{Autors et al.}

\author{G. Gogoberidze}
\affil{Institute of Theoretical Physics,
Ilia State University, 3 ave. Cholokashvili, Tbilisi 0162,
Georgia}
\and
\author{Y. M. Voitenko}
\affil{3 Solar-Terrestrial Centre of Excellence, Space Physics Division,
Belgian Institute for Space Aeronomy, Ringlaan-3-Avenue Circulaire,
B-1180 Brussels, Belgium}



\begin{abstract}
We study incompressible imbalanced magnetohydrodynamic turbulence in the
presence of background velocity shears. Using scaling arguments, we show
that the turbulent cascade is significantly accelerated when the background
velocity shear is stronger than the velocity shears in the subdominant Alfv%
\'{e}n waves at the injection scale. The spectral transport is then
controlled by the background shear rather than the turbulent shears and the
Tchen spectrum with spectral index $-1$ is formed. This spectrum extends
from the injection scale to the scale of the spectral break where the subdominant
wave shear becomes equal to the background shear. The estimated spectral
breaks and power spectra are in good agreement with those observed in the
fast solar wind. The proposed mechanism can contribute to enhanced turbulent
cascades and modified $-1$ spectra observed in the fast solar wind with
strong velocity shears. This mechanism can also operate in many other
astrophysical environments where turbulence develops on top of non-uniform
plasma flows.
\end{abstract}



\section{Introduction}

It is long known that the shear flows are important ingredient of the solar
wind dynamics. \citet{C68} was first who suggested that the solar wind is a
turbulently evolving medium. He noticed that velocity-shear driven
instabilities could produce power spectra of magnetic and velocity
fluctuations observed in the solar wind. He also suggested that the
dissipation of the turbulence at high wave numbers could account for the
anomalously high proton temperature observed in the solar wind at 1 AU.

The observations of the fast solar wind fluctuations \citep{BC13} show that below
the ion-cyclotron frequency spectrum of the fluctuations consist of two intervals.
Below the spacecraft-frame frequency $f_b \approx 10^{-3}$ Hz, which is usually referred as energy containing range, the spectral slope is close to $-1$, while for higher frequencies the Kolmogorov spectrum is observed (this range is called inertial range). It is widely agreed that the formation of the Kolmogorov spectrum in the inertial range is related to the active turbulent cascade, as originally proposed by \citet{C68}, whereas the origin of the spectrum observed in the energy containing range is not entirely clear yet [see, e.g., \citet{BC13} for a recent review]. First explanation was proposed by \citet{MG86}. These authors suggest that the observed spectrum results from the superposition of uncorrelated samples of solar surface turbulence. Alternative possibility suggests that the formation of the spectrum can be related to the coronal dynamics \citep{MEaO7}.

The viewpoint that velocity-shear driven
instabilities could produce power spectra of magnetic and velocity
fluctuations observed in the solar wind has two major shortcomings. Firstly, \citet{BD71} noticed
that Alfv\'{e}nic fluctuations in the fast flows of the solar wind are
strongly imbalanced - the power of the Alfv\'{e}n waves traveling outward
from the sun is significantly larger than the power of inward propagating
Alfv\'{e}n waves and it is difficult to explain how the shear-driven
instabilities can produce this asymmetry. Secondly, as
mentioned by \citet{BEa78}, the Kelvin-Helmholtz instability cannot produce
observed large-scale fluctuations. For these reasons it is widely accepted
nowadays \citep{BC13} that the dominant, outward traveling Alfv\'{e}n waves
are mainly generated near the Sun below the Alfv\'{e}nic critical point, as has
been originally proposed by \citet{BD71}.

On the other hand, even if generated, the inward waves can not propagate
above the Alfv\'{e}nic critical point and should have the local origin.
Moreover, analysis of Helios and Voyager data \citep{REa87} showed that
fluctuations in the solar wind become less imbalanced with increasing
distance from the sun. \citet{REa87} also found that the regions of strong
shear are associated with a rapid evolution from the purely Alfv\'{e}nic
state to a more balanced state with accelerated turbulent cascade. These
observations are puzzling in view of long known result of \citet{P64} that
the Kelvin-Helmholtz instability is inefficient in the solar wind. The same
result, strong enhancement of the turbulence cascade in strong shear flows,
has been confirmed later in numerical simulations \citep{GR99}. Therefore,
although enhancement of the turbulent cascade by the shear flows in seen
both in the solar wind observations and numerical simulations, many aspects
of its physics still remain unclear.

From the theory of neutral fluid turbulence it is long known \citep{T54,H}
that the strong background shears can significantly affect turbulent
dynamics. Namely, if the background velocity shear exceeds the velocity
shears in turbulent fluctuations, then the distortion of fluctuations driven
by the background shear dominates over nonlinear interactions. This leads to
the enhancement of the turbulence cascade rate and formation of so-called
Tchen spectrum $E(k)\sim k^{-1}$ (here $k$ is a wave number and $E(k)$ is
one dimensional spectrum of the fluctuations).

In this paper we consider strong incompressible imbalanced Alfv\'{e}nic
turbulence in the presence of background shear flows. By means of scaling
analysis we show that, similarly to the fluid turbulence, the strong shear
flow can significantly increase the energy cascade rate, resulting in the
formation of the Tchen-like spectrum. This is especially crucial for the
dominant Alfv\'{e}n waves, because their evolution driven by the subdominant
component is naturally weak because of the weak subdominant waves. Our
analysis shows that this mechanism can explain strong enhancements of
turbulent dynamics observed in the solar-wind shear flows.

The paper is organized as follows. Existing models of imbalanced
magnetohydrodynamical (MHD) turbulence are reviewed in Sec. 2. Phenomenology
of the Tchen model of strong imbalanced MHD turbulence in the presence of
strong velocity shear is developed in Sec. 3. Application of the obtained
results to the solar wind turbulence is discussed is Sec. 4 and conclusions
are given in Sec. 5.

\section{Existing Models of Anisotropic Imbalanced MHD Turbulence}

We consider incompressible MHD turbulence in the presence of the background
magnetic field $\mathbf{B}_{0}$. The Els\"{a}sser variables
\begin{equation}
\mathbf{w}^{\pm }=\mathbf{v}\pm \mathbf{b}/\sqrt{4\pi \rho },  \label{eq:0}
\end{equation}%
representing eigenfunctions of counter propagating Alfv\'{e}n waves, are
considered as the fundamental variables most useful to study MHD turbulence %
\citep{DMV83,B}. In equation (\ref{eq:0}) $\rho $ is the mass density, $\mathbf{v}$
and $\mathbf{b}$ are velocity and magnetic field fluctuations respectively.
The dynamics of the Els\"{a}sser variables is governed by the incompressible
MHD equations
\begin{equation}
\left( \frac{\partial }{\partial t}\mp \mathbf{V}_{A}\cdot \mathbf{\nabla }%
\right) \mathbf{w}^{\pm }+(\mathbf{w}^{\mp }\cdot \mathbf{\nabla })\mathbf{w}%
^{\pm }+\mathbf{\nabla }p=0.  \label{eq:01}
\end{equation}%
Here $p$ is the total (hydrodynamic plus magnetic) pressure and $\mathbf{V}%
_{A}\equiv \mathbf{B_{0}}/\sqrt{4\pi \rho }$ is the Alfv\'{e}n velocity. In
equations (\ref{eq:01}) we have neglected viscous and resistive dissipative
terms, which become important on smaller scales.

Alfv\'{e}n waves represent exact solutions of the ideal incompressible MHD
equations. This means that if in equations (\ref{eq:01}), say, $\mathbf{w}%
^{-}$ is zero initially, than $\mathbf{w}^{+}=\mathbf{w}^{+}(x,y,z-V_{A}t)$
is a nonlinear solution of arbitrary form. \citet{I63} and \citet{K65}
realized that due to this property, the MHD turbulence can be described as
nonlinear interactions of oppositely propagating Alfv\'{e}n wave packets.
The first model of MHD turbulence developed by \citet{I63} and \citet{K65}
assumed that the turbulence is isotropic. However, the mean magnetic field
has a strong effect on the turbulence, in contrast to the mean flow in the
hydrodynamic turbulence, which can be eliminated by the Galilean
transformation. The anisotropy of MHD turbulence had been already seen in
very early numerical simulations \citep{SMM83}.

A theory of anisotropic balanced (under balanced we mean turbulence with
equal energy of counter-propagating Alfv\'{e}n waves) MHD turbulence was
proposed by \citet{GS95}. This model implies that the dynamics of turbulence
is dominated by the perpendicular cascade with respect to the mean magnetic
field whereas the parallel size of turbulent 'eddies' (wave packets) is
determined by the critical balance condition. For wave packets with
characteristic parallel length scales $\Lambda ^{\pm }=\Lambda \sim
1/k_{\parallel }$ and perpendicular length scale $\lambda ^{\pm }=\lambda
\sim 1/k_{\perp }$, this condition implies that the characteristic time
scale of wave packet collision $\Lambda /V_{A}$ is equal to the
characteristic time scale of the energy cascade $t_{cas}\sim \lambda
/w_{\lambda }$, where $w_{\lambda }$ is characteristic value of the Elsasser variables at scale $\lambda$. As a result one arrives at Kolmogorov-like phenomenology
with $w_{\lambda }\sim \lambda ^{1/3}$. Equivalently, for 1-dimensional
perpendicular energy spectrum $E(k_{\perp })$ we have $E(k_{\perp })\sim
k_{\perp }^{-5/3}.$

In the case of imbalanced MHD turbulence situation becomes more complicated.
Assuming local turbulence, and noting that for Alfv\'{e}n waves Elsasser
fields $w_{\lambda }^{\pm }$ are perpendicular to the mean magnetic field,
it can be readily estimated that the nonlinear terms $(\mathbf{w}^{\mp
}\cdot \mathbf{\nabla })\mathbf{w}^{\pm }$ are of the order $\sim w_{\lambda
}^{+}w_{\lambda }^{-}/\lambda $. Therefore, the straining rates for $%
w_{\lambda }^{\pm }$ are \citep{LGSO7,CEaO9}
\begin{equation}
\omega _{sh}^{\pm }\sim \frac{w_{\lambda }^{\mp }}{\lambda }.  \label{eq:02}
\end{equation}%
If typical parallel length scale of colliding wave packets is $\Lambda $,
then characteristic timescale of their collision $\tau_{col}$ can be estimated as
\begin{equation}
\tau _{col}\sim \frac{\Lambda }{V_{A}}.  \label{eq:03}
\end{equation}
Note that  if to packets of size $\Lambda$ are counter-propagating with speed $V_A$, then collision time $\tau _{col}=\Lambda/(2V_A)$, but because we perform scaling analysis, this factor of 2 is ignored similar to other studies  \citep{LGSO7,CEaO9}.

We assume that $w_{\lambda }^{+}$ is the dominant component ($w_{\lambda
}^{+}\geq w_{\lambda }^{-}$). Dynamics of the turbulence depends on the
dimensionless parameter
\begin{equation}
\chi ^{+}=\tau _{col}\omega _{sh}^{+}\sim \frac{w_{\lambda }^{+}\Lambda }{%
V_{A}\lambda }.  \label{eq:04}
\end{equation}%
If $\chi ^{+}\gtrsim 1$, then subdominant wave packet is cascaded to smaller
scale during one collision and we have the strong turbulence. Then for the
energy cascade rate of the subdominant component we have
\begin{equation}
\varepsilon ^{-}\sim (w_{\lambda }^{-})^{2}\omega _{sh}\sim \frac{%
(w_{\lambda }^{-})^{2}w_{\lambda }^{+}}{\lambda }.  \label{eq:04a}
\end{equation}
This does not imply that the dominant wave packet is also cascaded during
one collision.

Regarding the cascade of the dominant waves, various models give different
predictions. Here we shortly consider main features and predictions of
several recent models of anisotropic imbalanced MHD turbulence. According to
the model developed by \citet{LGSO7} the straining rate imposed by the
subdominant waves on dominant ones, $w_{\lambda }^{-}/\lambda $, is imposed
coherently over a time $\lambda /w_{\lambda }^{-}$ and therefore cascade
time for the dominant waves is
\begin{equation}
\tau ^{+}\sim \frac{\lambda }{w_{\lambda }^{-}}.  \label{eq:05}
\end{equation}%
For the energy cascade rate of the dominant waves this equation gives
\begin{equation}
\varepsilon ^{+}\sim \frac{(w_{\lambda }^{+})^{2}}{\tau ^{+}}\sim \frac{%
(w_{\lambda }^{+})^{2}w_{\lambda }^{-}}{\lambda }.  \label{eq:06}
\end{equation}

According to the model developed by \citet{CO8}, the strainings of the
dominant waves by the subdominant ones are summed up randomly. This
assumption makes cascade of the dominant waves weaker:
\begin{equation}
\varepsilon ^{+}\sim \frac{w_{\lambda }^{+}\left( w_{\lambda }^{-}\right)
^{2}}{\lambda }.  \label{eq:07}
\end{equation}

Yet another model of strong imbalanced MHD turbulence was developed by %
\citet{BLO8}. The key feature of this model is that the turbulent
fluctuations of the dominant component cascade nonlocally, from $k_{1\perp }$
to significantly larger $k_{2\perp }$ where $k_{2z\left( -\right)
}=k_{1z\left( +\right) }$. As a result, the subdominant waves become more
anisotropic than the dominant waves. This model predicts the cascade rate of
dominant waves between the cascade rates predicted by two other models
(equations \ref{eq:06} and \ref{eq:07}), but there is no simple analytical
expression for this cascade rate.

\section{Tchen spectrum of MHD turbulence}

\citet{T54} was the first who recognized that the strong background shear
can significantly affect the energy cascade rate and statistical properties
of the hydrodynamic turbulence. In literature there exist several ways to
obtain the Tchen spectrum, including spectral energy budget analysis %
\citep{T54}, Heisenberg's eddy viscosity model \citep{KEa12} and scaling
analysis \citep{PEa86}.

Consider turbulent fluctuations of neutral fluid with characteristic
excitation scale $\lambda _{f}$ and amplitude $u_{f}$ imposed in the mean
flow with strong velocity shear, $S\equiv dV_{0}/dx\gg u_{f}/\lambda _{f}$.
Then the distortion of a turbulent eddy by the background flow is stronger
than the distortion by the turbulent flows (nonlinear interaction with other
eddies). The main effect of the sheared mean flow is stretching the eddies
along the flow, which in the wave number space is equivalent to the
increasing perpendicular (with respect to the mean flow) wave number.
Consequently, the background shear flow transfers the energy to higher
wave numbers faster than the nonlinear interactions.

If the fluctuations can be treated at outer scales as quasi-isotropic, then
at some scale $\lambda $ where the mean shear is greater that the inverse
eddy turnover time $v/\lambda $, the effective cascade timescale shortens
and becomes equal $\tau _{cas}\sim 1/S$ (although it has to be noted that
nonlinear interactions are still necessary to ensure decorrelation of
fluctuations and isotropic redistribution of fluctuation energy). If the
energy cascade rate is denoted by $\varepsilon $, then from equation $%
\varepsilon \sim v_{\lambda }^{2}/\tau _{cas}$ we have
\begin{equation}
v_{\lambda }\sim \sqrt{\frac{\varepsilon }{S}}.  \label{eq:08a}
\end{equation}%
For one dimensional energy spectrum $E(k)\sim v_{\lambda }^{2}/k$ this gives
\begin{equation}
E(k)\sim \frac{\varepsilon }{Sk}.  \label{eq:08}
\end{equation}%
Therefore, Tchen's model predicts that at relatively large scales, where the
shear imposed by the turbulent fluctuations is still weaker then the mean
flow shear, the energy spectrum should be inversely proportional to the
wave number, $E(k)\sim k^{-1}$. When $k$ increases, the shear associated with
the turbulent eddies $s_{\lambda }\sim kv_{\lambda }$ also increases and
starting from the wave number where $s_{\lambda }=S$ the turbulence is
expected to follow Kolmogorov's phenomenology. There is significant evidence
supporting Tchen spectrum both in boundary layer experiments and the
atmospheric boundary layer measurements [see, e.g., \citet{CEa13} and
references therein].

Here we develop an analogue of the Tchen phenomenology for the MHD
turbulence. Consider incompressible imbalanced MHD turbulence is the
presence of the background magnetic field $\mathbf{{B}_{0}\parallel z}$ and
background shear flow $\mathbf{{V}_{0}=}$ $\mathbf{(}0,0,Sx\mathbf{)}$.
Linear dynamics of MHD waves in such a flow have been studied by %
\citet{GEaO4}. Along with other phenomena (such as possibility of
over-reflection and mutual transformation of different MHD modes), one of
the main effects produced by the velocity shear is distortion of waves. In
the wave number space it is equivalent to the linear variation in time of the
perpendicular wave number, $k_{x}(t)=$ $k_{x}-Sk_{\parallel}t$. Similarly to the
hydrodynamic case, this is equivalent to the spectral transfer of energy in
the perpendicular wave number space. Therefore, with strong velocity shear
one can expect an enhancement of the cascade rate and formation of the
Tchen-type spectrum in the MHD turbulence.

Here we consider the strongly imbalanced turbulence, the reason for which is
twofold. First, the turbulence in the fast solar wind is strongly
imbalanced, and there is plenty of in-situ observations to compare with our
theoretical predictions. Second, in the imbalanced turbulence the cascade
rate of the dominant component is reduced significantly because of the low
amplitudes of subdominant waves responsible for the spectral transport in
the dominant component. Consequently, even relatively weak background shear
can strongly accelerate cascade in the dominant component.

Let us assume that the turbulence is excited isotropically at the (injection) outer scale $%
\lambda _{o}$ with the characteristic amplitudes of dominant and subdominant
components $w_{o}^{+}$ and $w_{o}^{-}$, respectively. Suppose that the
background velocity shear is moderately strong, exceeding velocity shears in
the subdominant component, but still smaller than the shears in the dominant
component:
\begin{equation}
\frac{w_{o}^{+}}{\lambda _{o}}>S>\frac{w_{o}^{-}}{\lambda _{o}}.
\label{eq:09}
\end{equation}%
In this case the cascade of subdominant waves is not significantly affected
by the background shear and the spectral flux is still given by equation (%
\ref{eq:04a}),
\begin{equation}
\varepsilon ^{-}\sim \frac{(w_{o}^{-})^{2}w_{o}^{+}}{\lambda _{o}}.
\label{eq:10}
\end{equation}

On the contrary, the strainings of dominant waves by the background shear
exceed the strainings imposed by the subdominant waves. Then, as in the
Tchen fluid model, the cascade time for dominant waves is effectively
shortened to $\tau _{cas}^{+}\sim 1/S$ and the cascade rate is accelerated
to $\gamma _{cas}^{+}\sim $ $1/\tau _{cas}^{+}\sim $ $S$. In terms of this
new cascade rate, the spectral flux in the wave number space at $k_{\perp
}\sim 1/\lambda $ is given by
\begin{equation}
\varepsilon ^{+}\sim {(w_{k}^{+})^{2}S}.  \label{eq:11}
\end{equation}%
Because of energy conservation, $\varepsilon ^{+}$ is constant and all terms
in this expression are $k$-independent, which results in the following
one-dimensional wave number spectrum of energy:
\begin{equation}
E^{+}(k)\sim \frac{{(w_{k}^{+})^{2}}}{k_{\perp }}\sim \frac{\varepsilon ^{+}%
}{S}k_{\perp }^{-1}.  \label{eq:12}
\end{equation}


The relative strength of the cascades generated by the background and
turbulent velocity shears can be conveniently described by the critical
parameter
\begin{equation}
\eta _{\lambda }\equiv \frac{S\lambda }{w_{\lambda }^{-}}.  \label{eq:12a}
\end{equation}%
The cascade is dominated by the background shear and the $-1$ spectrum (\ref%
{eq:12}) is formed at scales where $\eta _{\lambda }>1$. The turbulent
shears dominate at $\eta _{\lambda }<1$ forming the $-5/3$ spectrum.

Equations (\ref{eq:09}-\ref{eq:12a}) represent our model of the imbalanced
MHD turbulence modified by the velocity shear. If the imbalanced MHD
turbulence follows phenomenology by \citet{LGSO7}, then formation of Tchen's
spectrum is expected if the background shear is strong enough in sense of
equation (\ref{eq:09}), i.e. when the cascade rate due to background shear ($%
\gamma _{cas}^{+}\sim S$) is larger than the cascade rate due to the
turbulent shears at the injection scale ($\gamma _{o}\sim w_{o}^{-}/\lambda
_{o}$):
\begin{equation}
\eta _{\lambda _{o}}>1.  \label{eq:13}
\end{equation}
In the cases where the turbulence follows phenomenology by \citet{CO8} with
a weaker cascade of dominant waves, the Tchen spectrum can be formed by the
proportionally smaller background shear (then the critical parameter $\eta
_{\lambda }$ should be modified correspondingly).

As the cascade generated by the background shear proceeds to smaller scales,
the Tchen cascade rate $S$ remain the same. On the contrary, the strainings
imposed by the turbulent eddies become progressively stronger because of the
stronger velocity gradients in the small-scale eddies. Then $\eta _{\lambda }
$ decreases below $\eta _{\lambda _{o}}$ and the Tchen-type cascade
eventually arrives to the spectral break
\begin{equation}
\lambda _{b}=\frac{w_{b}^{-}}{S},  \label{eq:13a}
\end{equation}%
where $\eta _{\lambda _{b}}=1$, i.e. the background and turbulent shears
become the same. The Tchen wave number spectrum $\sim k_{\perp }^{-1}$ is
formed at scales $\lambda _{o}>$ $\lambda >$ $\lambda _{b}$, whereas the
strongly turbulent spectrum $\sim k_{\perp }^{-5/3}$ is formed at smaller
scales $\lambda <$ $\lambda _{b}$.

\section{Application to the solar wind turbulence}

Recent studies based on in-situ observations have revieled that the
fast-slow solar wind interface has two parts: a smooth "boundary layer"
surrounding the fast wind, and a sharper "discontinuity" between the slow
and intermediate solar winds \citep{SEaO5}. A relatively strong velocity
shear was observed by Ulysses over its first orbit in the transition area
between the fast and slow solar winds at $13^{\circ }-20^{\circ }$
latitudes. The data analysis \citep{MEa98} showed that the boundary layer
separating two winds consists of two regions, the first one with the width $%
l_{1}\approx 2\times 10^{7}$ $\mathrm{km}$ and velocity difference $\Delta
V_{1}\approx 200$ $\mathrm{km/s}$ and the second one with $l_{2}\approx
8\times 10^{7}$ $\mathrm{km}$ and $\Delta V_{2}\approx 100$ $\mathrm{km/s}$.

The spectral brake between the "energy containing range" and the "inertial
range" occurs at the spacecraft-frame frequency $f_{b}\approx \times 10^{-3}
$ $\mathrm{Hz}$ \citep{TEa15,BC13}. The corresponding break scale. It is well know
that the power of inward propagating Alfv\'{e}n waves in the fast solar wind
streams is about one order of magnitude lower than the power of outward
waves [see, e.g., \citet{WEa11} and references therein]. As the typical
values in the fast solar wind we take $w_{b}^{-}\sim 7$ $\mathrm{km/s}$ for
the subdominant wave amplitude at the scale $\lambda _{b}$ %
\citep{WEa11,GEa12} and $V_{sw}\approx 600$ $\mathrm{km/s}$ for the solar
wind speed \citep{BC13}.

Noting that $\lambda _{b}=V_{sw}/f_{b}$ and $S=\Delta V_{1}/\Delta l_{1}$,
with observed numerical values our model predicts
\begin{equation}
f_{b}=\frac{V_{sw}\Delta V_{1}}{\Delta l_{1}w_{b}^{-}}\approx 1.2\times
10^{-3}~\mathrm{Hz.}  \label{eq:15}
\end{equation}
As we see performed rough estimate gives the value which is the same order of magnitude as the observed spectral brake frequency. Below $f_{b}$ our model predicts the Tchen spectrum $\sim k_{\perp }^{-1}$. Although we do not claim that all observed $%
\sim k_{\perp }^{-1}$ spectra are generated by our mechanism, the
correspondence between the model and observations is good enough to motivate
further observational studies. In particular, as the break wave number $%
k_{\perp b}$ between $\sim k_{\perp }^{-1}$ and $\sim k_{\perp }^{-5/3}$
spectra is proportional to the background shear $S$, the presence of
positive correlation between $k_{\perp b}$ and $S$ in various data sets of
fast solar wind streams would strongly support our mechanism.

As it is known \citep{BC13} the $-1$ spectrum is not observed in the slow solar wind. Therefore another interesting direction of further research is to study weather this phenomenon is related to the absence of strong velocity shear in the slow solar wind.

\section{Conclusions}

We developed a semi-phenomenological model of incompressible imbalanced MHD
turbulence in the presence of sheared background flows. Our results can be
summarized as follows:\

1) The Tchen-type spectrum $\sim k_{\perp }^{-1}$ can be generated by the
background velocity shear exceeding the shears of the subdominant Alfv\'{e}n
waves at the injection scale $\lambda _{o}$.

2) The $k_{\perp }^{-1}$ spectrum breaks down at the scale $\lambda _{b}$
given by (\ref{eq:13a}), where the turbulent shears of the subdominant
component become as strong as the background shear. The $k_{\perp }^{-1}$
spectrum extends from $\lambda _{o}$ to $\lambda _{b}$.

3) At smaller scales, $\lambda <\lambda _{b}$, the Kolmogorov $k_{\perp
}^{-5/3}$ spectrum is formed by the turbulent velocity shears.

It is long known, but still unexplained, that in the fast solar wind streams
the spectral index of turbulent fluctuations at large scales is close to $-1$
and the spectral break frequency is close to $ 10^{-3}$ $\mathrm{Hz}$
(see e.g. \citet{M91}). These observations are compatible with the mechanism
we propose here, which motivates its future verification by observations.

Our model can be applied to other astrophysical environments with strong
velocity shears, like astrophysical jets and supernova explosions.


\acknowledgments
 This work has been supported by Shota Rustaveli National Science
Foundation grant FR/51/6-300/14.

\end{document}